\documentclass[12pt,onecolumn,prb,showpacs]{revtex4}
\usepackage{graphicx,color}
\usepackage{amsmath}
\usepackage{longtable}
\begin{document}

\title{Self-diffusion in strongly coupled Yukawa systems (complex plasmas)}

\author{Sergey A. Khrapak,$^{1,2,}\footnote{Electronic mail: skhrapak@mpe.mpg.de}$ Olga S. Vaulina,$^2$ and Gregor E. Morfill$^1$}
\date{\today}
\affiliation{ $^1$Max-Planck-Institut f\"ur extraterrestrische Physik, Postfach 1312, Giesenbachstr., 85741 Garching,
Germany \\$^2$Joint Institute for High Temperatures, 125412 Moscow, Russia}

\begin{abstract}
We show that the idea of mapping between the Newtonian and Brownian diffusivities proposed and tested on a class of particle systems interacting via soft and ultra-soft potentials (IPL, Gaussian core, Hertzian, and effective star-polymer) by Pond {\it et al}., [Soft Matter {\bf 7}, 9859 (2011)] is also applicable to the Yukawa (screened Coulomb) interaction. Some of the implications of this result with respect to self-diffusion in strongly coupled complex (dusty) plasmas are discussed.
\end{abstract}

\maketitle

\section{Introduction}

Recently Pond {\it at al}.~\cite{Pond} proposed a simple idea for mapping between long-time self-diffusion coefficients obtained from molecular and Brownian dynamics simulations. Their primary interest was on particles interacting via soft inverse-power-law (IPL) potentials as well as ultrasoft Gaussian core, Hertzian and effective star-polymer interactions. The latter three models have nowadays received considerable attention in the field of soft condensed matter. A simple heuristic expression relating Newtonian and Brownian diffusivities has been demonstrated to describe reasonably well extensive simulation data for the class of systems investigated.~\cite{Pond}

The main purpose of the present work is to check whether the proposed mapping is adequate for complex (dusty) plasmas, which have been recently recognized as new class of soft matter -- the ``plasma state of soft matter''.~\cite{MorfillRMP,ChaudhuriSM} Complex plasmas consist of weakly ionized gas (conventional plasma) and highly charged macroscopic (dust) particles.~\cite{Vladimirov,Book,FortovPR} Highly charged particles interact with each other electrically. In the first approximation this interaction can be modeled by the Debye-H\"{u}ckel (screened Coulomb or Yukawa) potential, although its actual shape can be considerably more complicated, especially at large interparticle separations.~\cite{Book,FortovPR,Daugherty,KhrapakPRL2008,KhrapakCPP} The electrical interaction energy can often be remarkably high as compared to the particle kinetic energy and, therefore, complex plasmas can be viewed as classical systems of individually visible strongly interacting particles. They can be used (complimentary to other soft matter systems like e.g. colloids and granular medium) to investigate a broad range of important fundamental processes, such as phase transitions and self-organization, phase separation, rheology, waves, transport, and other.

The focus of the present study is on the diffusion-related properties of complex plasmas. We have to admit that diffusion in Yukawa systems has been already investigated in a number of papers. For instance, molecular dynamics technique has been employed in Refs.~\onlinecite{Robbins,Ohta} to simulate self-diffusion coefficients of one-component Yukawa systems in the fluid phase in a wide range of thermodynamical parameters. Brownian dynamics technique has been used in Ref.~\onlinecite{Lowen} to obtain the diffusion coefficient of an overdamped Yukawa fluid near the freezing point. Comparison between molecular dynamics and Brownian dynamics diffusivities in a system of polydisperse Yukawa particles near a glass transition towards an amorphous solid has been performed in Ref.~\onlinecite{Lowen1991}. A Langevin-type approach has been used to study diffusion in the fluid phase of Yukawa systems over a broad range of damping strength.~\cite{Vaulina2001,Vaulina2002,VauVla} These latter studies are particularly relevant to complex plasmas since the coupling strength between the particles and the plasma background (i.e. damping, which is usually dominated by the neutral gas component) can be varied in a very broad range. As a result, complex plasmas can be in principle ``engineered'' as essentially a one component system (when damping is vanishingly small and the Newtonian dynamics is dominated by the interparticle interactions), or as an overdamped system of Brownian interacting particles (when damping from the background medium dominates dynamical phenomena).~\cite{KhrapakPRE2004} This explains why the possibility to map between molecular dynamics (Newtonian) and overdamped (Brownian) diffusivities represents considerable interest in the context of complex plasmas. In the rest of this paper we use the available numerical results  to demonstrate the applicability of the mapping proposed in Ref.~\onlinecite{Pond} to strongly coupled Yukawa systems. Moreover, we take a next step and suggest an expression for the diffusion coefficient, which is applicable to strongly coupled Yukawa systems with an arbitrary damping strength.

\section{Normalizations}

The Yukawa pair potential can be written as
\begin{equation}
U(r)=\epsilon(\lambda/r)\exp(-r/\lambda),
\end{equation}
where $\epsilon$ is the energy scale, $\lambda$ is the screening length, and $r$ is the distance between two particles. For charged particles immersed in a plasma the energy scale is related to the particle charge $Q$ ($\epsilon= Q^2/\lambda$) and the screening comes from the equilibrium redistribution of plasma electrons and ions in the vicinity of the particle  ($\lambda=\lambda_{\rm D}$, where $\lambda_{D}$ is the Debye radius). The Yukawa system in thermodynamical equilibrium can be characterized by two dimensionless parameters. One natural choice is the reduced temperature $T_*=T/\epsilon$ and density $\rho_*=\rho\lambda^3$. In the field of complex (dusty) plasmas, however, it is more usual to use the {\it coupling parameter} $\Gamma=Q^2/(T\Delta)$ and the {\it screening parameter} $\kappa=\Delta/\lambda$, where $\Delta=\rho^{-1/3}$ is the mean interparticle distance. Note that the fluid-solid phase transition in Yukawa systems can be to a good accuracy described by $\Gamma\simeq 106\exp(\kappa)(1+\kappa+\frac{1}{2}\kappa^2)^{-1}$, provided $\kappa$ is not too large.~\cite{VaulinaJETP2000} We will denote the value of the coupling parameter corresponding to this phase transition as $\Gamma_{\rm M}$, where the subscript ``M'' refers to melting (similarly, $T_{\rm M}$ corresponds to the temperature at melting).~\cite{Note}

In the following we will be dealing with the diffusion coefficients of 3D Yukawa systems in the fluid phase only. The actual (dimensional) long-time diffusion coefficient is denoted by $D$, its values in the limiting regimes of molecular (Newtonian) dynamics and Brownian (overdamped) dynamics are denoted by $D_{\rm MD}$ and $D_{\rm BD}$ respectively, following the notation of Ref.~\onlinecite{Pond}. For Newtonian diffusion without damping we employ Rosenfeld's normalization~\cite{Rosenfeld} and use the reduced diffusion coefficient in the form $D_{\rm RMD}=D_{\rm MD}\rho^{1/3}\sqrt{M/T}$, where $M$ is the particle mass, and the temperature $T$ is in energy units. For Brownian diffusion the natural choice of reduced diffusivity is $D_{\rm BD}/D_0$, where $D_0$ is the value of $D_{\rm BD}$ in the dilute limit (or, equivalently, in the absence of interparticle interactions).

\section{Mapping}

A heuristic expression describing mapping between Newtonian and Brownian diffusivities proposed in Ref.~\onlinecite{Pond} is
\begin{equation}\label{mapping}
D_{\rm BD}/D_0 = 1 - (1+c_1D_{\rm RMD}+c_2D_{\rm RMD}^{3/2})^{-1},
\end{equation}
where $c_1=3.3176$ and $c_2=2.6645$ are numerical constants. It describes relatively well the numerical data for IPL interactions as well as for model complex fluids with soft star-polymer, Gaussian-core, and Hertzian interactions.~\cite{Pond}

\begin{figure}
\centering
\includegraphics[width= 8 cm]{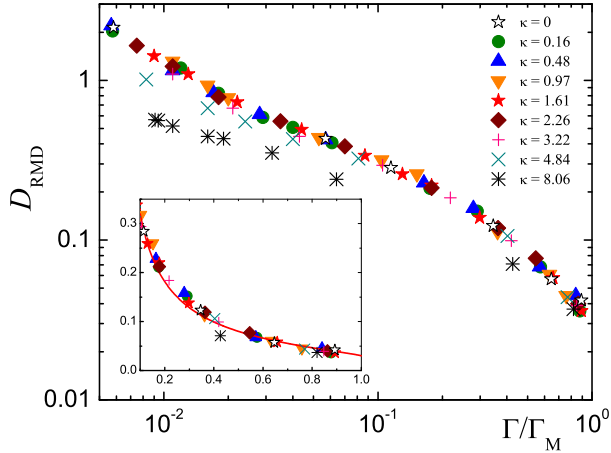}
\caption{Reduced diffusion coefficient $D_{\rm RMD}$ in the molecular dynamics (Newtonian) regime vs. the relative coupling strength $\Gamma/\Gamma_{\rm M}$. Numerical data are shown by symbols. The data for the Yukawa potential are taken from Ref.~\onlinecite{Ohta}. The data for the one-component-plasma (OCP) limit are from Ref.~\onlinecite{Hansen}. Inset shows the portion of the data corresponding to the strongly coupled regime. The solid curve is the fit using Eq.~(\ref{RMD}).}\label{Fig1}
\end{figure}

In order to verify whether this mapping is applicable to Yukawa interactions, let us first consider the available numerical data for the Newtonian limit. Figure~\ref{Fig1} shows the reduced diffusion coefficients $D_{\rm RMD}$ plotted against the ratio $\Gamma/\Gamma_{\rm M}=T_{\rm M}/T$, i.e., as a function of the distance from the melting line. Numerical data for Yukawa systems are taken from MD simulations by Ohta and Hamaguchi.~\cite{Ohta} As discussed in Ref.~\onlinecite{Ohta}, the numerical values of the diffusion coefficient may be subject to errors of up to about 10$\%$. The data for the limiting case ($\kappa=0$) of one-component-plasma (OCP) are those from MD simulations by Hansen {\it et al}.~\cite{Hansen} Figure~\ref{Fig1} demonstrates that the reduced diffusion coefficient $D_{\rm RMD}$ shows a quasi-universal behavior, except in the regime of sufficiently steep interaction ($\kappa\gtrsim 4$) and weak coupling, where systemmatic deviations (decrease in $D_{\rm RMD}$) are evident. In the strongly coupled regime not too far from the melting line (say, $0.1\Gamma\lesssim \Gamma\lesssim \Gamma_{\rm M}$) the data are scattered near a single curve (see inset in Fig.~\ref{Fig1}), which can be reasonably fitted by the function
\begin{equation}\label{RMD}
D_{\rm RMD}\simeq \alpha\left(T/T_{\rm M}-1\right)^{\beta}+\gamma,
\end{equation}
with the coefficients $\alpha\simeq 0.05$, $\beta\simeq 0.8$, and $\gamma\simeq 0.03$. The same functional form has been used in Ref.~\onlinecite{Ohta}, but with different coefficients, because of the different normalization for the diffusivity employed there.

\begin{figure}
\centering
\includegraphics[width= 8 cm]{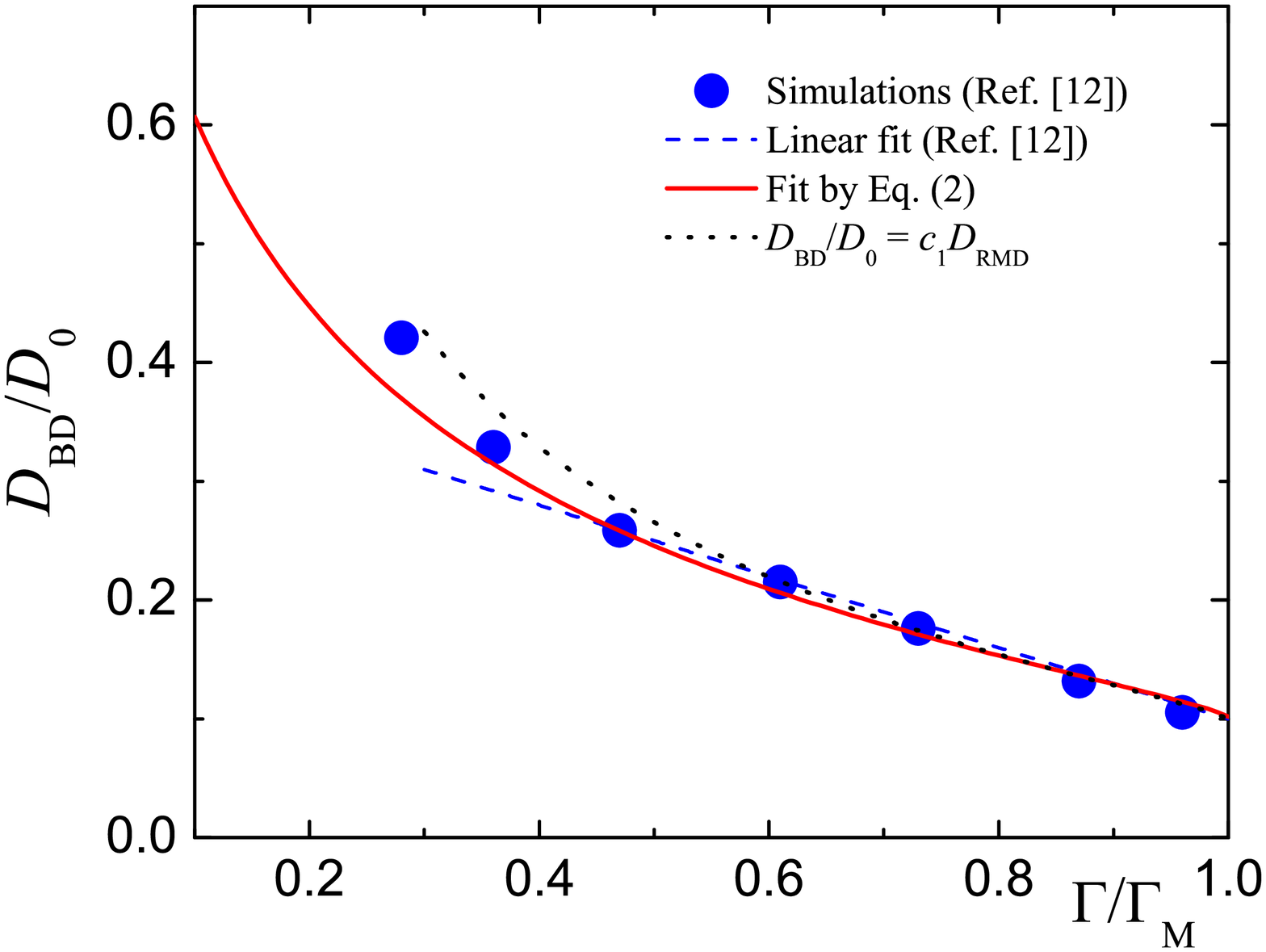}
\caption{Reduced diffusion coefficient $D_{\rm BD}/D_0$ in the overdamped (Brownian) regime vs. the relative coupling strength $\Gamma/\Gamma_{\rm M}$. Numerical data shown by symbols are from Ref.~\onlinecite{Vaulina2002}. The solid curve represents the mapping by Eq.~(\ref{mapping}). The dotted curve corresponds to its simplified version (\ref{proportionality}). The dashed line is the linear fit suggested in Ref.~\onlinecite{Vaulina2002} for the strongly coupled limit. All three curves and symbols almost fall on top of each other for $\Gamma\gtrsim 0.5\Gamma_{\rm M}$.}\label{Fig2}
\end{figure}

With the use of Eqs.~(\ref{mapping})-(\ref{RMD}) we can immediately calculate the dependence of the reduced Brownian diffusion coefficient $D_{\rm BD}/D_0$ on the ratio $\Gamma/\Gamma_{\rm M}$ (and $T_{\rm M}/T$). The results are shown in Fig.~\ref{Fig2} by the solid curve. Here we also show a simple linear dependence
\begin{equation}
D_{\rm BD}/D_0\simeq 0.3(1-T_{\rm M}/T)+0.1,
\end{equation}
proposed by Vaulina {\it et al}.~\cite{Vaulina2002} for strongly coupled Yukawa systems ($\Gamma\gtrsim 0.5\Gamma_{\rm M}$) in the limit of strong damping (dashed line) along with the numerical data from the same work corresponding to the highest (but finite) damping rate (circles). The agreement is convincing, which proves the applicability of the mapping (\ref{mapping}) to Yukawa systems in the strongly coupled regime.

\section{Discussion}

Let us discuss an important consequence of the applicability of the mapping (\ref{mapping}) to strongly coupled Yukawa systems. First of all note that on approaching the melting line, both $D_{\rm RMD}$ and $D_{\rm BD}$ become small (see Figs.~\ref{Fig1} and \ref{Fig2}). Hence, to the lowest order in $D_{\rm RMD}$ we have
\begin{equation}\label{proportionality}
D_{\rm BD}/D_0\simeq c_1 D_{\rm RMD},
\end{equation}
i.e. both reduced diffusivities are proportional to each other. As we have seen already, in a broad range of parameters corresponding to strong coupling, the value of $D_{\rm RMD}$ is governed by a single quantity -- the ratio of $\Gamma/\Gamma_{\rm M}$ (or $T_{\rm M}/T$). Let us denote ${\mathcal F}(\Gamma/\Gamma_{\rm M})=c_1D_{\rm RMD}$. Then the actual diffusion coefficient in the two limiting regimes considered can be presented as
\begin{equation}\label{temptation}
D=D_0\left\{\begin{array}{l}
\frac{\nu_{\rm fr}\Delta}{c_1}\sqrt{\frac{M}{T}}{\mathcal F}(\Gamma/\Gamma_{\rm M}) \qquad\quad ({\rm Newtonian}) \\[.2cm]
{\mathcal F}(\Gamma/\Gamma_{\rm M}), \qquad\qquad\qquad  ({\rm Brownian})
\end{array}
\right.
\end{equation}
where $\nu_{\rm fr}$ is the macroscopic friction rate associated with particle interaction with the surrounding medium (neutral gas in complex plasmas), so that the bare Brownian diffusion coefficient (no interaction) is $D_0=T/(M\nu_{\rm fr})$. The ``damping index'' $\xi=\nu_{\rm fr}\Delta\sqrt{M/T}$ naturally measures the damping strength. Physically, $\xi$ is the ratio between a typical interparticle spacing and the mean ballistic free path of the particles. For $\xi\ll 1$ the particle motion is essentially ballistic and this corresponds to the molecular dynamics (Newtonian) limit. Contrary, for $\xi\gg 1$ the particle motion is diffusive on the length scales considerably shorter than the interparticle distances. This corresponds to the overdamped (Brownian) limit. Examination of Eq.~(\ref{temptation}) immediately suggests that it is convenient (and not very unreasonable) to separate the effect of interparticle interactions and damping and write the actual diffusion coefficient in the form
\begin{equation}\label{decoupl}
D=D_0 {\mathcal F}(\Gamma/\Gamma_{\rm M}) {\mathcal G}(\xi),
\end{equation}
where the function ${\mathcal F}$ is responsible for the suppression of self-diffusion due to (strong) interparticle interactions, while ${\mathcal G}$ is the ``damping function'' describing the transition from the limit of Brownian dynamics [$\xi\gg1$ and ${\mathcal G}(\xi)\rightarrow 1$] to the limit of Newtonian dynamics [$\xi\ll 1$ and ${\mathcal G}(\xi)\rightarrow \xi/c_1$].

\begin{figure}
\centering
\includegraphics[width= 8 cm]{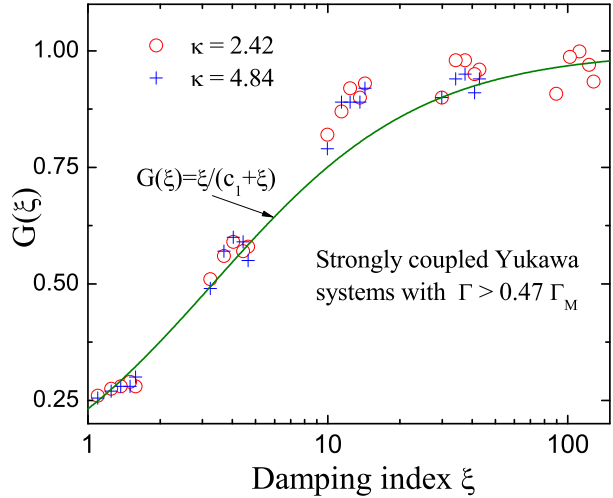}
\caption{The damping function ${\mathcal G}$ vs. the damping index $\xi=\nu_{\rm fr}\Delta\sqrt{M/T}$ for strongly coupled Yukawa systems. Symbols correspond to the numerical data from Ref.~\onlinecite{Vaulina2002}. The solid curve is a simple approximation given by Eq.~(\ref{approx}).}\label{Fig3}
\end{figure}

We use the numerical data from Ref.~\onlinecite{Vaulina2002} to check the plausibility of the scaling (\ref{decoupl}). These data were obtained for the two values of the screening parameter ($\kappa=2.42$ and $\kappa=4.84$) and various damping indexes $1\lesssim \xi\lesssim 100$. In figure \ref{Fig3} we plot the corresponding ratios $D/[D_0{\mathcal F}(\Gamma/\Gamma_{\rm M})]\equiv {\mathcal G}(\xi)$. Although the data points are somewhat scattered, they clearly tend to group near a certain curve. Thus, in the regime of strong coupling, Eq. (\ref{decoupl}) represents a reasonable approximation (we found it, however, less accurate for weaker damping). A simple function of the form
\begin{equation}\label{approx}
{\mathcal G}(\xi)=\frac{\xi}{c_1+\xi}
\end{equation}
is exact in the respective limiting cases and provides a reasonable description of the numerical results in the transitional regime. More statistics and higher data accuracy are apparently required to construct a better analytical expression for ${\mathcal G}(\xi)$.

\section{Concluding remarks}

We have demonstrated that the idea of mapping between the Newtonian and Brownian diffusivities proposed and tested on a class of soft repulsive potentials (IPL, Gaussian core, Hertzian, and effective star-polymer) in Ref.~\onlinecite{Pond}, also works relatively well for the Yukawa interaction potential. This can have important implications for the field of complex plasmas, where the coupling strength between the particles and surrounding background (typically dominated by neutral gas) can be varied in a broad range, so that both limiting regimes (as well as intermediate states) can be realized.

We have also seen that in a wide parameter regime the Newtonian diffusivity $D_{\rm RMD}$ of Yukawa systems exhibits universal behavior with respect to the reduced coupling strength, provided   the interaction is soft enough ($\kappa\lesssim 4$). An interesting issue to investigate is whether this or similar scaling holds for other soft and ultrasoft interactions, for which the universal excess entropy scaling~\cite{Rosenfeld} does not work.~\cite{Pond,Krekelberg}

We have proposed an expression for the diffusion coefficient applicable to an arbitrary damping efficiency in the system. In this expression the ratio of the actual diffusion coefficient to its bare Brownian value is given by a product of the two functions, one of which describes suppression of the self-diffusion due to interparticle interactions and depends only on the relative coupling strength, while the other -- the damping function -- is a convenient measure of the damping strength and is independent of the interaction details. Such an approach demonstrates reasonable accuracy when applied to Yukawa systems in the strongly coupled regime. Thus, it can be used to estimate the diffusion-related quantities in complex plasmas near the fluid-solid (crystallization) phase change.  The important question to be answered in future is whether this scaling is truly universal and/or how wide is the class of interaction for which it holds.

Finally, it would be interesting to look at the observed universalities from the perspective of the emerging principle of corresponding states for strongly coupled particle systems, which has been recently put forward in Ref.~\onlinecite{Principle}.


\begin{thebibliography}{99}
\bibitem{Note} We do not distinguish between freezing and melting since the miscibility gap is rather small in the regime of not too large $\kappa$ considered here.
\end{thebibliography}
\end{document}